\documentclass[aps,prl,preprint]{revtex4-1}
\usepackage{graphicx}
\usepackage{latexsym}
\usepackage{amsmath}
\usepackage{amssymb}
\begin{document}
\title{Diamond monohydride: The most stable three-dimensional hydrocarbon}
\author{M.V.Kondrin}
\email{mkondrin@hppi.troitsk.ru}
\author{V.V. Brazhkin}
\affiliation{Institute for High Pressure Physics RAS, 142190 Troitsk, Moscow, Russia}

\begin{abstract}
Most of hydrocarbons are either molecular structures or linear polymeric chains. Discovery of graphene and manufacturing of its monohydride -- graphane incite interest in search for three-dimensional hydrocarbon polymers. However up to now all hypothetical hydrocarbon lattices significantly lose in energy to stacked graphane sheets and solid benzene. We propose completely covalently bonded solid carbon monohydride whose density significantly exceeds one of its isomers (graphane, cubane, solid benzene). {\em Ab-initio} calculation demonstrates that the cohesion energy of this structure at least is not worse than the energy of graphane and benzene. In some aspect the crystal structure of hydrocarbon presented can be regarded as a sublattice of diamond, but with the symmetry of $P\overline{3}$ space group (lattice parameters  $a \approx  6.925$ \AA\ , $c \approx 12.830$ \AA\ ) and Z=42 formula units per unit cell. This structure (if synthesized) may turn out to be interesting to applications.
\end{abstract}

\maketitle
\section{Introduction}
Carbon is known to have a large number of modifications with various valence (2--4) and hybridization types. The likely cause for that is the small size of carbon's core electron shell and large number of outer electrons. So up to now there are a lot of known allotropic phases of carbon (and much more of hypothetical ones) with different dimensionality. Among the known phases  are quasi-0D molecular fullerite-like phases, 1D -- carbynes, 2D -- graphite and nanotubes, 3D -- diamond and its hexagonal counterpart lonesdaleite. In the experimentally attainable pressure range on P-T diagram the stability regions exist for only two phases -- graphite and diamond (whose cohesive energies are practically the same and differs only by 20 meV), all others are metastable with much lesser values of cohesive energies per atom. The existence of metastable carbon phases is due to strongly directed covalent bonds and respectively high activational energy barriers\cite{brazhkin:pu09}. 

When hydrogen atoms are included into these carbon carcasses the number of possible structures increases dramatically. There is nothing astonishing that the number of known organic compounds (not taking into account biological structures) is more than 30 millions. However among hydrocarbon subset of organic structures there is only one the most stable compound -- methane, all others are metastable\cite{brazhkin:pu06}. The large size of hydrocarbon molecules and the presence of saturated C-H bonds leads to certain restrictions caused by the steric interactions of hydrogen atoms which in general prevents formation of fully covalently bonded structures. As a result most of hydrocarbons crystallizes in molecular structures and only application of high pressures, strong catalyzers or technologically complicated processes allows one to obtain polymerized phases which as a rule are quasi-1D chains or strongly disordered mixture of amorphous/crystalline phases.

There is marked interest to organic compounds with 1:1 C:H stochiometry.  One can expect that the dimensionality of these phases can be compared with that of pure carbon allotropes. However, up to the recent times only 0D (molecular phases of benzene, styrol, acetylene, cubane/cuneane) and 1D were considered. Among linear hydrocarbon compounds one can mention well-known polyacytelene, polystyrol and  more complicated polymers of benzene, so-called nanothreads, which were proposed\cite{wen:jacs11} and synthesized\cite{fitzgibbons:nm14} recently. However among these phases the most stable one is benzene. 

Until the discovery of graphene the organic structures with higher dimensionality were proposed only occasionally\cite{sluiter:prb03,sofo:prb07}. However the successful synthesis of graphane\cite{elias:s09} (that is 2-dimensional hydrocarbon film where hydrogen atoms are bound to graphene plane and so induces sp$^3$-hybridization accompanied with  corresponding corrugation of the carbon plane) instigates interest in synthesis of hydrocarbon polymers in higher dimensions. The first move in this direction was the proposition of van-der-Waals bonded graphane sheets\cite{wen:pnas11} (that is graphane sheets stacked in direction normal to the plane; juxtaposition graphene-graphite is quite relevant in this regard) but almost in the same time the efforts of invention of hydrocarbon structures comletely covalently bounded in 3D took place. Up to now there were proposed at least four hydrocarbon structures of such a type\cite{lian:jcp13,he:jpcm13,lian:sr15}. Although they are more energetically favourable than cubane but all of them significantly loose in energy (more than 0.3 eV per CH group) to molecular benzene\cite{ciabini:prb05,ciabini:nm07} and stacked graphane sheets \cite{wen:pnas11}. Such a difference in energies are mostly due to significant deviation of bond lengths and angles from the most energetically favourable tetrahedral one (i.e. ideal sp$^3$ bond observed for example in diamond). 

In this paper we would like to present another hypothetical 3D allotrope of carbon monohydride (we will call it diamond monohydride/DMH) which can be energetically as viable as graphane. It will be shown that DMH can be regarded as sublattices of diamond structure so internal strain in it (as well as in graphane) is minimal and mostly caused by steric interactions of hydrogen atoms. Also we will demonstrate that the proposed 3D structure is the densest monohydride with expected numerical density of carbon atoms about one half of the diamond one.
\section{The problem}
It is remarkable that the problem of search for the least strained substructures of some known crystal structure can be posed in quite abstract way. Say we have some 3D structure (diamond lattice in our case) regarded as a 3D network. We want to exclude some nodes from it according to some rules thus providing enough room to put into this structure required number of guest atoms (hydrogen in our case).  In the case of carbon and hydrogen the restrictions are due to the respective length of C-C covalent bond (1.54 \AA\ in diamond-like structure) and C-H bond length ($\approx$1.0 \AA). Ideally, C-H bond should point along the line between empty and occupied cites thus exerting minimal strain in the lattice. However, each empty cite can have at most only one occupied cite in its neighborhood (otherwise the distance between hydrogen atoms on the bonds converging to the same empty node would be too short). From this immediately follows that the maximal numerical density (that is the density of occupied sites) of structures produced in such a way will be one half of the initial.  In the case of networks with maximal density one can also note certain congruency between the two sets consisting of occupied and empty nodes -- each of them has exactly one neighbor of other sort.

It should be noted that the two types of graphane sheets (named A and B in Ref.~\onlinecite{wen:pnas11}) can be built (in {\em gedunkenexperiment}) from diamond structure according to this routine, provided that  the nodes are removed along two differently oriented planes. The same is true for the more corrugated graphane sheet (named tri-cycled graphane) proposed in Ref.~\onlinecite{he:pssr12}. However in all these cases the contiguity of the initial structure isn't retained -- there are no covalent bonds between graphane sheets.

So our result demonstrates that in the case of diamond-like network there is at least one  example of the contiguous (that is completely covalently bonded) substructure with maximal density that satisfies all these conditions. However the questions whether this structure is the only one and what about less dense structures are still open. We also want to stress that our procedure is different from the one adopted in Refs.~\onlinecite{lian:jcp13,he:jpcm13,lian:sr15}, where authors start from sp$^2$ hybridized networks of benzene or pure carbon (with addition of some amount of hydrogen atoms in the latter case) and look for more energetically favorable configuration of atoms. In our case, on the other hand, the type of hybridization (sp$^3$) persists throughout the procedure.
\section{Methods}
For the search of this structure a number of programs from Bilbao crystallography server was used\cite{aroyo:zfk06,capillas:jac03} . For calculation of internal energy of DMH and its elastic moduli we used Quantum ESPRESSO software package\cite{gianozzi:jopcm09}. For density functional calculation Perdew-Burke-Ernzerhof exchange correlation method was applied with project augmented waves pseudopotentials for both carbon and hydrogen atoms with energy cutoff 70 Ry and charge-density cutoff 800 Ry. For integration over Brillouin zone unshifted $8 \times 8 \times 8$ Monkhorst-Pack grid was used. In the process of calculation relaxation of cell dimensions and ion  positions (with fixed initial symmetry) was done. Before the calculation of properties and elastic stability evaluation, the crystal lattices and atom positions are fully optimized until the residual force on every atom is less than 0.001 Ry/bohr and the residual stress is less than 0.5 kbar. For preparation of input files and visualization of output \verb|cif2cell|\cite{cif2cell} and \verb|Jmol|\cite{jmol} programs were actively used.
\section{Cohesive energy}
\begin{figure}
\includegraphics[width=0.8\columnwidth]{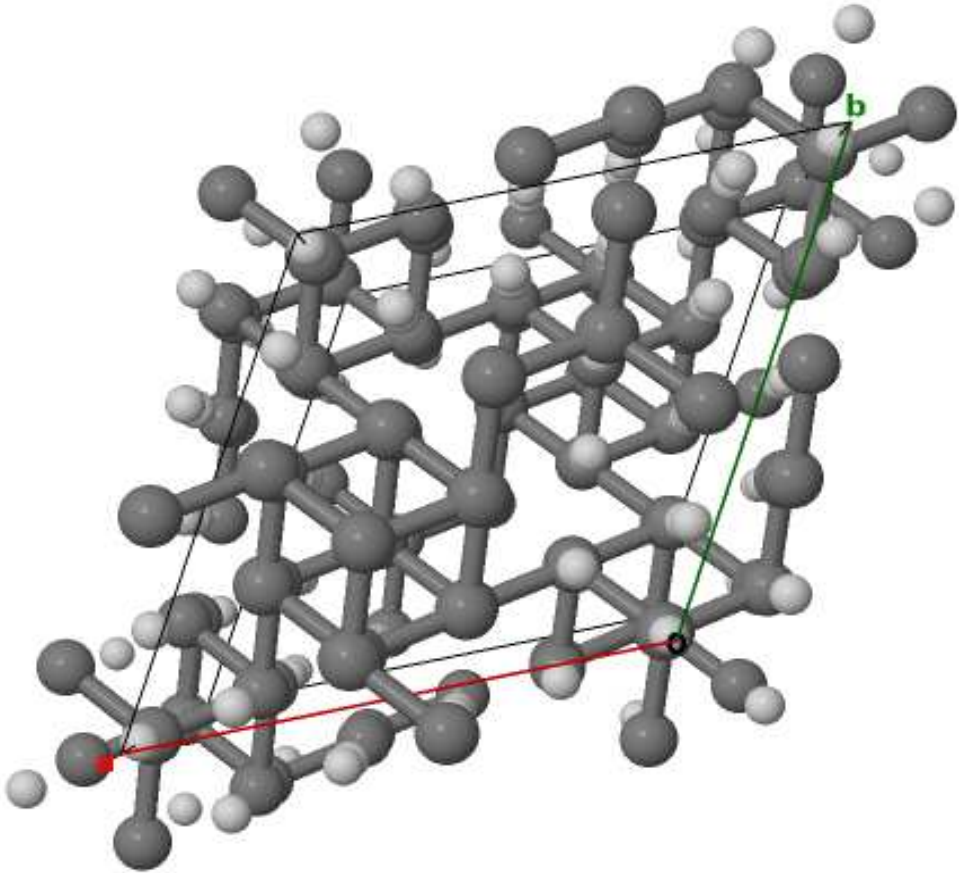}
\caption{Perspective view along hexagonal axis of $P\overline{3}$ sublattice of diamond structure. Light gray balls mark the positions of unoccupied nodes, gray balls and sticks designate the remnants of the original lattice.}
\label{st0}
\end{figure}

The resulting structure can be conveniently described by the sequence of virtual symmetry breaking operations. First, the rhombohedral ``distortion'' of diamond-like crystal along one of the space diagonals leads to the space group transformation $Fd\overline{3}m \xrightarrow{(8,1)} R\overline{3}$ (the numbers above arrows designate {\em translationengleiche} and {\em klassengleiche} subgroup indices $t,k$). Then the formation  in two dimensions along the basal plane of 7-fold supercell (it can be thought of as a central hexagonal cell surrounded by the six others) leads to the symmetry breaking  $R\overline{3} \xrightarrow{(1,7)}  R\overline{3}$. And the last transformation is the doubling of the period along hexagonal axis  which produces the transition $R\overline{3} \xrightarrow{(1,6)} P\overline{3}$. The result of composition of all these three transitions is  splitting of $8a$ Wyckoff position of the diamond-like structure  to the two $2c$, four $2d$ and twelve $6g$ sets of $P\overline{3}$ space group so that each of them can be evenly distributed among occupied and unoccupied nodes (the resulting structure is shown in Fig.~\ref{st0}). From this consideration immediately follows that the resulting conventional unit cell (obtuse hexagonal) has dimensions $a \approx 6.67$ \AA\ , $c \approx 12.36$ \AA\ with 42 formula units (occupied-unoccupied site pairs) in it. 

\begin{figure}
\includegraphics[width=0.8\columnwidth]{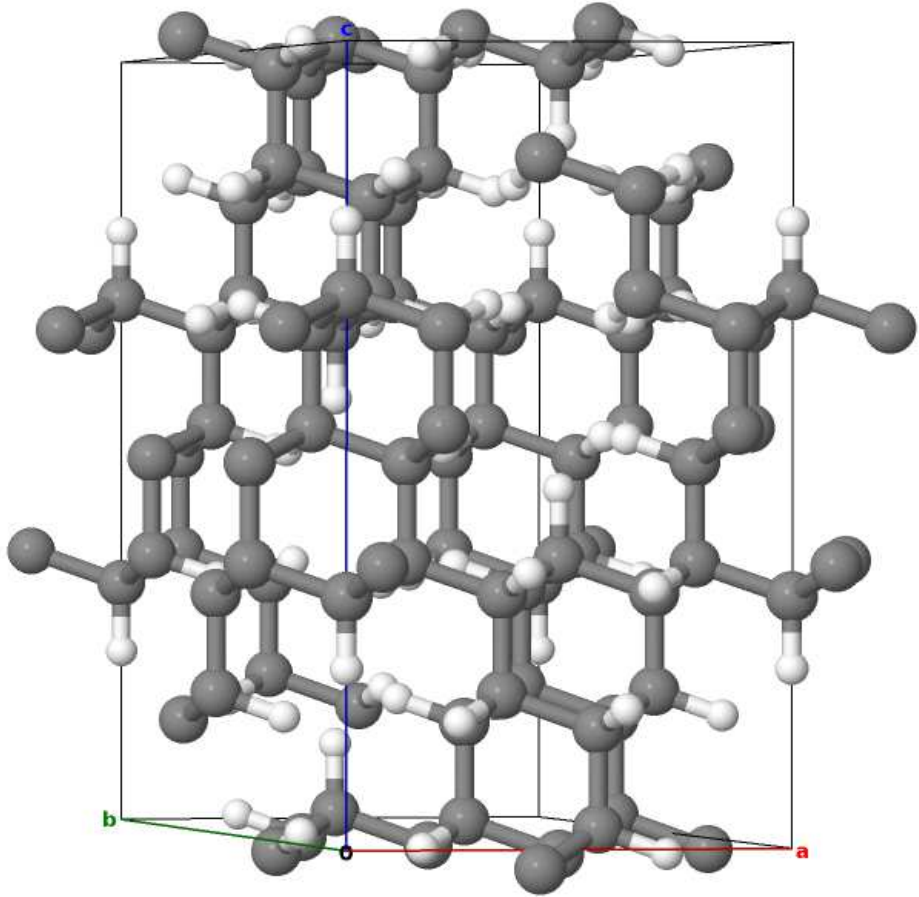}
\caption{Perspective view along basal plane of the unrelaxed structure. Light gray and dark gray balls designate hydrogen and carbon atoms respectively.}
\label{st1}
\end{figure}

As the initial guess for structure optimization  hydrogen atoms are put into this structure 1 \AA\ apart of occupied sites in the direction pointing to the closest unoccupied one. As it was demonstrated above, in every case there is the only one such a site (see Fig.~\ref{st1}). After optimization of cell dimensions and ion positions in the same $P\overline{3}$ symmetry the structure distorts a little but its internal energy is still very close to that of graphane I \cite{wen:pnas11} and benzene. In our computation we actually obtain that the energy at the ambient pressure of benzene and DMH per CH are practically the same, though DMH is slightly more favorable. The cohesive energy of benzene and graphane I is about 100-150 meV (per CH) lower  than that of DMH. In our opinion the result can be significantly influenced by the van-der-Waals interaction which is not taken into account in our calculation, so the energy of all these compound can be considered practically the same with the error of about 50 meV per CH.

\begin{figure}
\includegraphics[width=0.8\columnwidth]{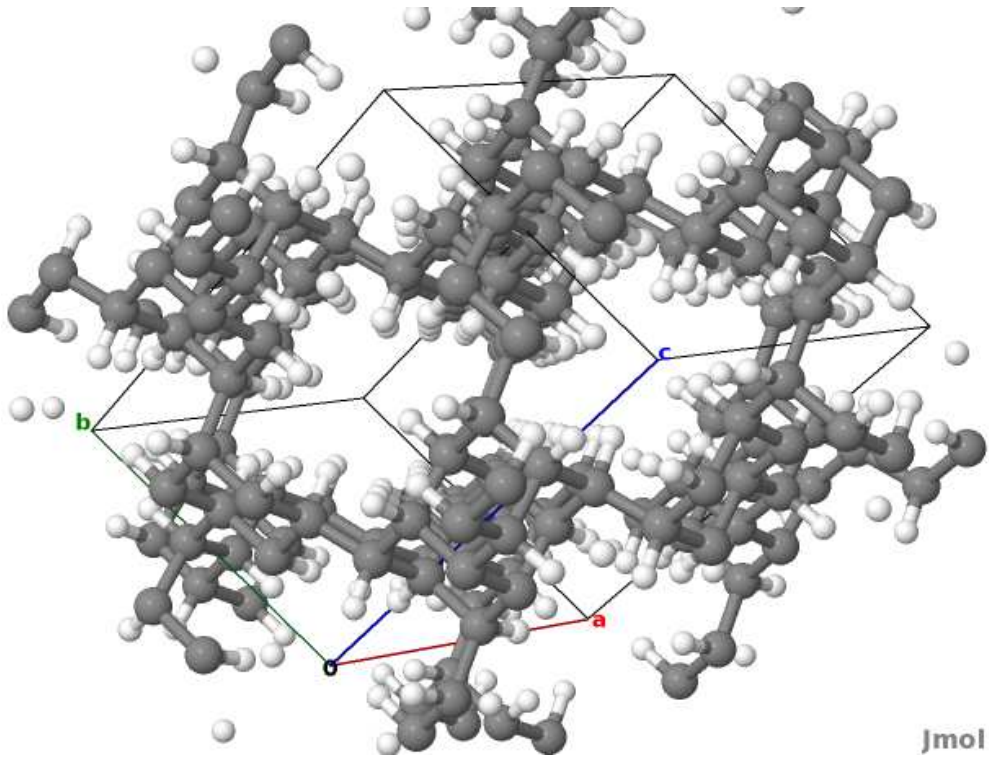}
\caption{Relaxed DMH structure. Perspective view along pores.}
\label{st2}
\end{figure}

\begin{figure}
\includegraphics[width=0.8\columnwidth]{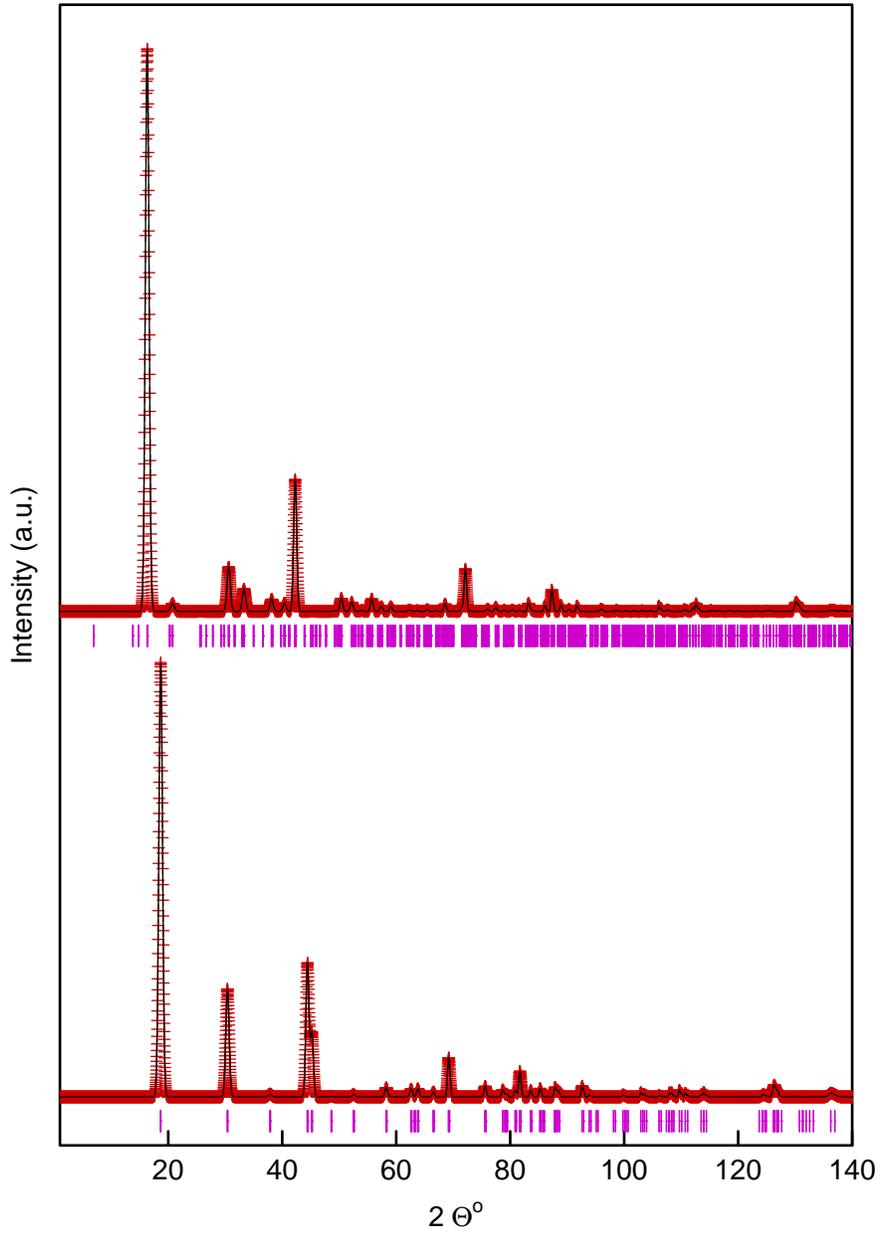}
\caption{Simulated XRD pattern of graphane III\cite{wen:pnas11,lian:sr15} (below) and DMH for the $K^{Cu}_{\alpha}$ wavelength.}
\label{xray}
\end{figure}

Electronic bands calculation demonstrates that DMH is a good insulator with the energy gap of 4.5 eV, so in this regard it is comparable with graphanes. The calculated mass density is 1.7 g/cm$^3$, it is $\approx$10\% less than that of unrelaxed structure due to corresponding increase of the unit cell, but nonetheless at ambient pressure it is higher than that of graphanes\cite{wen:pnas11} ($\approx1.5-1.6$ g/cm$^3$) and cubane\cite{yildirim:prl97} ($\approx$1.3 g/cm$^3$). The minimal distances between atoms in this structure are $C-C=1.56$, $C-H=1.10$ and $H\cdot \cdot H=2.05$ \AA\ . Detailed crystallographic information about DMH is presented in Table~\ref{supplement}, simulated X-Ray pattern is shown in Fig.~\ref{xray}. There is marked  difference in XRD pattern of the proposed hydrocarbon and the 3D structures suggested before -- in our case the diffraction pattern starts from significantly lower angles (16.5$^o$ in Fig.~\ref{xray} vs. $\approx 23^o$ in Ref.~\cite{lian:sr15}). Another interesting feature of this structure is the presence of quite spacious ``tubes'' filled with hydrogen piercing the bulk of the structure. The look along one of such tubes is shown in Fig.~\ref{st2}. Similar pores are observed in other 3D hydrocarbons \cite{lian:sr15} but they are significantly  more narrow. 

\begin{table*}[h]
\caption{Coordinates of atoms in the unit cell}
\label{supplement}
\begin{tabular}{lllll}
\toprule%
 Atom & site & X & Y & Z \\\hline
C &   2d &     0.333333333 &   0.666666667 &   0.652991873 \\
C &   2c &     0.000000000 &   0.000000000 &   0.680338216 \\
C &   2d &     0.666666667 &   0.333333333 &   0.986322101 \\
C &   6g &     0.566693935 &   0.835558168 &   0.812300515 \\
C &   6g &     0.766635541 &   0.831106449 &   0.521032851 \\
C &   6g &     0.900051526 &   0.502242228 &   0.145638177 \\
C &   6g &     0.569642262 &   0.843085045 &   0.691800814 \\
C &   6g &     0.763690856 &   0.823581186 &   0.641531155 \\
C &   6g &     0.902971125 &   0.509747015 &   0.025135604 \\
H &   2d &     0.333333333 &   0.666666667 &   0.567542759 \\
H &   2c &    -0.000000000 &  -0.000000000 &   0.765789504 \\
H &   2d &     0.666666667 &   0.333333333 &   0.900880204 \\
H &   6g &     0.007808087 &   0.413550060 &   0.330108651 \\
H &   6g &     0.325523639 &   0.253115590 &   1.003226790 \\
H &   6g &     0.341140932 &   0.080216668 &   0.663440226 \\
H &   6g &     0.110353010 &   0.399959709 &   0.491271272 \\
H &   6g &     0.222982801 &   0.266707443 &   0.842060946 \\
H &   6g &     0.443681631 &   0.066615362 &   0.824605790 \\\toprule
\end{tabular}
\end{table*}
\section{Elastic stability}
The demonstration of viability of hypothetical crystal structure requires not only the calculation of its energy but also the proof of its dynamical stability, that is the absence of very soft phonon modes\cite{sandro:zkri05}. Due to the large unit cell (84 atoms in the unit cell) we didn't calculate phonon modes of DMH. Still we can show that this structure is elastically (or statically) stable, so it satisfies Born's stability criterion. 

In general Born's criterion states that the energy of any crystal structure should increase under strain. So it can be fast and roughly checked by calculation of the energy of the same structure under pressure. In the case of DMH we fully relaxed it at pressure 2 GPa and  found that its energy increases. Corresponding bulk modulus $B$ calculated using the formulas:
\begin{align}
B=-V(\partial P /\partial V)_T,\nonumber\\
P=-(\partial E/ \partial V)_T\nonumber
\end{align} 
 with discrete differences substituted for partial derivatives yields the value $B \approx 80 $ GPa. It is more than 5 times lower than that of diamond, but taking into account that DMH was ``obtained'' by removing some of the covalent bonds from diamond-like structure there is nothing unusual in this result. However it is larger than the elastic moduli of graphanes ($15-30$ GPa\cite{wen:pnas11}) but less than the that of 3D-hydrocarbon structures proposed before ($180-200$ GPa \cite{lian:sr15}). So this might indicate a correlation between empty space present in the crystal lattices of isomeric compounds and their respective hardness. 

However more careful consideration of Born's stability requires that 4-order elastic tensor $c$ has to be positively definite (so that {\em any} combination of strains leads to increase of energy). In case of $\overline{3}$ point group this is equivalent to the series of inequalities between 7 independent elastic constants\cite{mouhat:prb14}:
\begin{align}
c_{11}>|c_{12}|,c_{44}>0,\nonumber\\
c_{13}^2<\frac{1}{2}c_{33}(c_{11}+c_{12}),\nonumber\\
c_{14}^2+c_{15}^2<\frac{1}{2}c_{44}(c_{11}-c_{12})\nonumber
\end{align}
Here we apply Voigt notation and replace $3 \times 3$ symmetrical strain matrix $\delta$ with 6 dimensional vector $\varepsilon$ (accompanied by the corresponding decreasing of elastic tensor's order):
\begin{equation}
\delta=
\begin{pmatrix}
\varepsilon_1 & \varepsilon_6 / 2 & \varepsilon_5/2\\
\cdot & \varepsilon_2 & \varepsilon_4/2 \\
\cdot & \cdot & \varepsilon_3
\end{pmatrix}
\nonumber
\end{equation}

For calculation of individual elastic constants the unit cell $R$ (regarded as matrix consisting of coordinates of unit vectors) was deformed applying some type of strain $\delta$ with amplitude 0.05 according to equation $R'=R\cdot \delta$. The internal energy was obtained after relaxation of ion positions in this transformed unit cell\cite{catti:aca89}. After that the corresponding elastic constant was derived from the energy difference between initial and strained structures using the equation: $\Delta E = \sum \frac{1}{2}c_{ij}\varepsilon_j^2$. The similar results can be obtained from the stress matrix/vector $\sigma=c \cdot \varepsilon$ produced in the course of computation. However in both these cases one should pay attention to the lowering of crystal lattice symmetry caused by the strain so in the most general case one should take into account all 21 independent elastic moduli. Nonetheless our calculation for DMH structure yields the values  consistent in 5\% margin: $c_{11}=140, c_{33}=250, c_{44}=80, c_{12}=90, c_{13}=45, c_{14}=15, c_{15}=-15$ GPa. It can be easily checked that these elastic moduli satisfy Born's criteria. The averaging of elastic constants yields bulk modulus value $B = 90 \pm 15$ GPa (which is fairly coincides with the one obtained from DFT calculations directly) and shear modulus $G=60 \pm 15$ GPa.
\section{Conclusions} 
In this paper we have demonstrated the viability of covalently bonded hydrocarbon structure with $P\overline{3}$ symmetry and 42 CH pairs in the unit cell. DFT calculations suggest that the lattice parameters of this structure are $a \approx  6.925$ \AA\ , $c \approx 12.830$ \AA\ and the density $\rho = 1.7$ g/cm$^3$. This structure should be insulating with energy gap 4.5 eV. Calculation of its mechanical properties demonstrates its elastic stability. The values of bulk $B=80$ GPa and shear  $G=60$ GPa moduli allow one to classify this material as quite hard one. Since the calculated internal energy of DMH was found to be at least not worse than the energies of  benzene and graphane, we suppose that in practice the stability region of this structure might exist somewhere on the P-T phase diagram of C:H mixture. Due to its large bandgap and nanoporous structure, this compound may turn out to be interesting as a material for optoelectronic and biological applications or as solid hydrogen storage element.
\begin{acknowledgments}
The authors acknowledge  financial support from the RSF (Grant No. 14-22-00093) and thanks Y. B. Lebed, A. G. Lyapin and A. R. Oganov for helpful discussions. M. V. K. thanks Quantum ESPRESSO developers and users for their patient and prompt answers in \verb|pw_forum@pwscf.org| mailing list.
\end{acknowledgments}
\bibliography{ch15}
\end{document}